**PAPER • OPEN ACCESS**

# Spinal Muscle Atrophy Disease Modelling as Bayesian Network

To cite this article: Mohammed Ezzat Helal *et al* 2021 *J. Phys.: Conf. Ser.* **2128** 012015

View the article online for updates and enhancements.

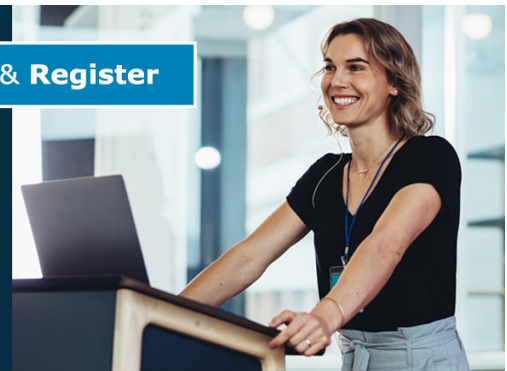

# Spinal Muscle Atrophy Disease Modelling as Bayesian Network

**Mohammed Ezzat Helal[1], Manal Ezzat Helal[2] and Professor Sherif Fadel Fahmy[1]**

**Abstract.** We investigate the molecular gene expressions studies and public databases for disease modelling using Probabilistic Graphical Models and Bayesian Inference. A case study on Spinal Muscle Atrophy Genome-Wide Association Study results is modelled and analyzed. The genes up and down-regulated in two stages of the disease development are linked to prior knowledge published in the public domain and co-expressions network is created and analyzed. The Molecular Pathways triggered by these genes are identified. The Bayesian inference posteriors distributions are estimated using a variational analytical algorithm and a Markov chain Monte Carlo sampling algorithm. Assumptions, limitations and possible future work are concluded.

## 1   Introduction

Human cells undergo cell activities that vary over time based on genetic makeup, environmental factors, drugs interference, disease development and recovery. These variations can be computationally modelled using dynamic networks over time and analysed to identify therapeutic opportunities, causal analysis, and various parameter estimate analyses. PGM has been applied to model human diseases as identified in the literature review section. However, there are inconsistencies in the results reporting of published papers and how they are stored in public databases and their availability for the wider community to apply and learn new models from them.

To the best of our knowledge, many publicly available analysis algorithms and methods use toy examples or manually analysed datasets and encourage others to re-apply on new datasets. This work aims to identify a pipeline of analysis steps to achieve a disease model that can be used to add new evidence to create a dynamic ongoing model as more studies are reported. This paper attempts to study the Spinal Muscle Atrophy (SMA) disease, identify relevant Genome-Wide Association Study (GWAS) and build Probabilistic Graphical Models (PGM) to capture the uncertainty in gene expressions and estimate future expressions, identifying assumptions, limitations and challenges. The model can be enhanced to generalise from and resume the proposed future work to overcome the identified limitations.

## 2   LITERATURE REVIEW

### 2.1 Probabilistic Graphical Models

PGM captures the joint distribution between observed random variables as the nodes of a graph/network, in which edges denote dependency between the variables. When the graph is directed and acyclic (DAG), a Bayesian Network (BN) is created, while an undirected graph creates a Markov Random Fields (MRF). Once a model is defined from a dataset, further analysis can be applied to it. The possible analysis is 1) learning the parameters of the network, such as the conditional probability distributions;

---

[1] Computer Engineering Department, AASTMT, Cairo, Egypt, fahmy@aast.edu
[2] School of Physics, Engineering and Computer Science, University of Hertfordshire, Hatfield, UK,
m.helal@herts.ac.uk

2) learning the structure by selecting a model that best fits the data to a probability distribution with the highest Bayesian score or the Bayesian information criterion (BIC), and 3) building an inference engine to respond to probabilistic queries [1]. PGMs may include latent variables that are linked to observed variables. Multilevel Hierarchical Latent Class models are useful to capture mixture models and enable clustering, dimensionality reduction, and latent causal inference [2].

The simplest application of PGMs has been in modelling hereditary features between family members [3, p. 9]. DNA sequences in the living cells are inherited strings that are divided into chromosomes (23 in humans from each parent). A basic DNA interaction cycle is defined as DNA is divided into genes that transcribe mRNA that translates proteins and enzymes that also interact together and with their cell environment and affect their future expressions. Gene products and various cell metabolites form networks of interactions and affect pathways or functions by inhibition or activation. This cycle continues with time interacting with various environmental factors that constantly change the cells' contents (by increasing or decreasing) of these metabolites. These interactions can be captured in a hierarchical PGM, in which vertices are of different metabolite types (each type has its hierarchical level) and edges are the various captured interactions. Such PGM would be very high dimensional. Studies focus on modelling a particular disease (phenotype) and the set of genes involved in its most significant interactions [4].

## 2.2 PGMs for GWAS Data

The work in [2], [5] identifies various modelling approaches applying PGM to model diseases GWAS output compared to the Population Association Studies (PAS). PAS quantifies how case and control cases differ in their allelic frequencies identifying heredity. These kinds of studies enable PGMs to model pedigrees (individuals with specific phenotype groups encoded by their genotype) to model the heredity in the meiosis process to query simple linkage analysis and detection of quantitative trait locus. Linkage disequilibrium mapping then identifies the precise location of the gene responsible for the observed phenotype. Another pedigree PGM model uses random variables as any of the allelic types from both parents. Phenotypic PGM models on the other hand add an extra node to the individual onto which the phenotype prevails [2].

PGMs are also modelled using Single Nucleotide Polymorphism (SNP) as the random variables capturing the dependencies between SNPs. Various models are discussed in [5] to query the SNP-phenotype association. The GWAS produces a vast amount of data compared to PAS and previous PGM models will be intractable if applied to the complete dataset. Selecting the most significant SNPs to reduce the model's complexity led to losing the benefits of the genome-wide analysis as the preselection of what to include might be incorrect. Scalable solutions include variable length Markov Models (VLMMs) that groups genotypes into clusters or patterns. Other models study the multilocus SNP-disease association by limiting the physical distance between SNPs and/or by using decomposable graphs such as GraphMiner. Graphical Lasso enables sparse SNP-SNP dependencies without distance restrictions. This is besides the gene-gene interactions (epistasis) that is modelled in DASSO-MB (Detection of ASSOciations using Markov Banket) models, MDR (Multifactor Dimensionality Reduction) and BEAM (Bayesian Epistasis Association Mapping). Another modelling approach applies data integration of genetics, gene expressions and proteomics to capture complete biological processes [4].

GWAS studies produce gene expression counts by producing fragments that are probabilistically assigned to specific genes. Chapter 5 in [6] explains how to create a PGM model for the GWAS data using Bayesian inference to estimate the gene expression parameters for one or more genes, in one or more samples. The model discusses the different required probability distributions of all parameters involved in the experiment, such as the gene expression count, random error, depth of the experiment and variability within groups. The model uses Markov Chain Monte Carlo (MCMC) algorithms implemented in Stan to derive the full posterior distribution from the defined conditional dependence, priors, and likelihood functions.

The work in [7] explains how to reverse engineer the Gene Regulatory Network (GRN) from GWAS gene counts using Hierarchical Bayesian Models including known data from candidate cis-regulatory modules (CRMs) and Transcription Factors (TF) regulating the genes in the experiment to infer the binary (on or off) and activity (activation or depression or 0) assuming Normal Distribution for all variables for simplicity. The latter method uses Gibbs sampling to infer for each gene the posterior probability of the CRMs associated with it.

The work in [8] uses the Bayes Factors of Covariance Structures (BFCS) to infer gene-gene causal relationships from GWAS data. They aimed to derive the posterior probability of the causal relationship of a triplet structure connecting pairs of traits (pair of genes corresponding to the traits) and a genetic marker as one gene as a cause that regulates the other gene as an effect. This local search is proven faster and more effective than other methods, as it considers all structures at once and is inherently parallel.

### 2.3 Spinal Muscle Atrophy (SMA) Modelling

PGMs can model known information about diseases. Spinal Muscle Atrophy (SMA) is characterised by being a neurodegenerative disorder caused by the mutation in chromosome 5 at position q13.2 identifying the gene as SMN. SMN was found to be duplicated as SMN1 and SMN2 with 5 nucleotides difference between them including a C to T mutation in exon 7. The normal SMN1 gene usually produces 100% SMN protein, while the mutated SMN2 creates an exonic splicing suppressor (ESS) that leads to skipping of exon 7 during splicing and production of truncated, non-functional SMN protein. The SMN2 gene can still produce around 10% of full-length mRNA. The copy number of SMN2 has been correlated with the disease severity, with SMA Type I and II commonly have two or three copies of SMN2 creating 20:30% of needed SMN. SMA Type III has four copies creating 40% of the needed SMN protein. 50% of SMN protein is identified as sufficient to have normal function [9].

SMA has been modelled using PGM in [10] using hierarchical nodal structures and links between model parameters. The model was created by expert knowledge and observational studies, without publishing their results. An SMA GWAS study in a fruit fly model (Drosophila) was conducted in [11] collecting gene expressions variations between control (normal) and case (SMN mutants) in the second and third instar larval stages. The motility defects usually prevail between these two stages. The gene expressions have been collected from brain and muscle tissues in fold change with a cut-off of 1.5. The experimental setup, equipment, and standardisation parameters are further detailed in [11].
Small n large p problem prevails in these experiments, as only two cell types and two disease development stages were tested, for many gene expressions. Moreover, not all genes were expressed in all stages, creating a sparse matrix. These gene expressions as the observed variables can be used to construct PGM to infer the distribution of latent variables representing the biological processes or functions or pathways these genes are involved in [12].

## 3    METHODOLOGY

### 3.1 Pre-processing

The GWAS SMA study in [11] produced 3158 genes' expression in the fold change values in two different time intervals and two different cell types (Brain and Muscle) in different files divided by up and down-regulated genes. The first pre-processing step was collecting the data per gene, per stage and per tissue type, and consider downregulation as -ve values of the change fold. This step produced an n × p data matrix X, comprising n observations, which are cell types and time step or stage in the experiment, for p molecular variables, which are genes.

The second step was to find the Entrez Ref Sequence ID through python Bio package code, which searches for the gene name in the nuccore database. Preference was given to IDs starting with "NM_" then "XM_" prefix to favour protein-coding transcribed mRNA over predicted ones and other entries containing the query name as produced from the GWAS. This collection of IDs was generated from a sequence of searches from the most preferred RefSeq and mRNA and animal organisms to the most unrestricted search, then compared to select the most relevant as possible [13]. The difficulty in this step was in the availability of many variants and many different types of the same organism. This is because many studies do not report full details about the organism, or the specific gene variant. This reduced the genes to 3149 genes with RefSeq ID [11].

The third pre-processing step was to do Pathway Enrichments to reduce dimensionality. The genes were enriched using the Reactome knowledgebase [14] to identify the pathways they are involved in. Only the genes regulating some pathways were used in the analysis. This step was processed using the KNIME platform pipeline for gene expression enrichments [15]. The workflow started by identifying the Fold Change of the counts of genes in several samples using the edgeR Bioconductor tool to identify the significantly differentially expressed genes. This step was already done in the produced files of the SMA GWAS study in [11]. The significantly expressed genes as illustrated in Figure 1 were then clustered hierarchically based on their expression pattern. This step produced a heatmap as illustrated in Figure 2 and a dendrogram as illustrated in Figure 3 containing 23 significant genes that have common expression patterns. A pathway enrichment analysis step identified the pathways the significantly expressed genes are involved in as illustrated in Figure 4. The most significant pathway was the Xenobiotics on the top of Figure 4, which is a metabolism activity that handle foreign compound in the organism such as drug or poison. The second pathway affects the growth of the child which is expected in SMA disease.

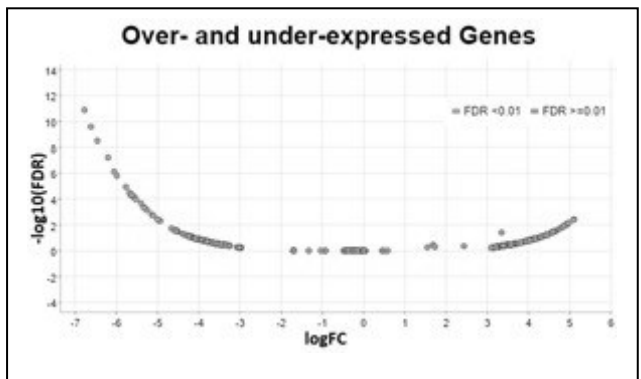

*Figure 1: Over- and under-expressed Genes*

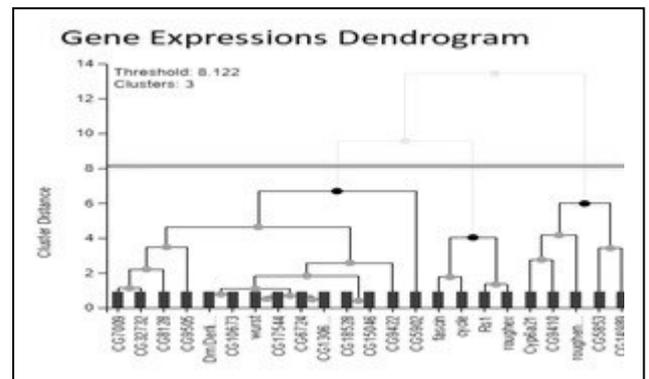

*Figure 3: Dendrogram for significantly expressed genes based on similar expression patterns*

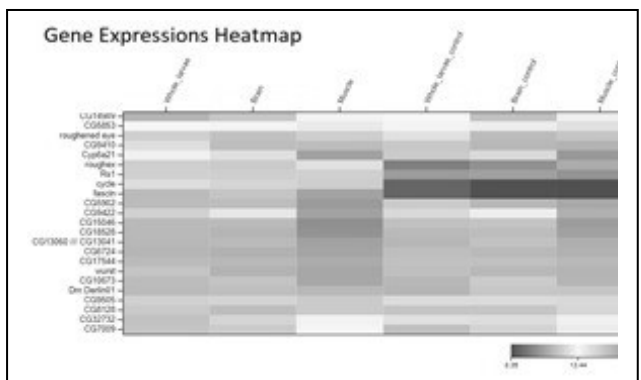

*Figure 2: Heatmap for significantly expressed genes across samples from Whole Larves, Brain, and Muscle in stage three as compared to the Control in stage two*

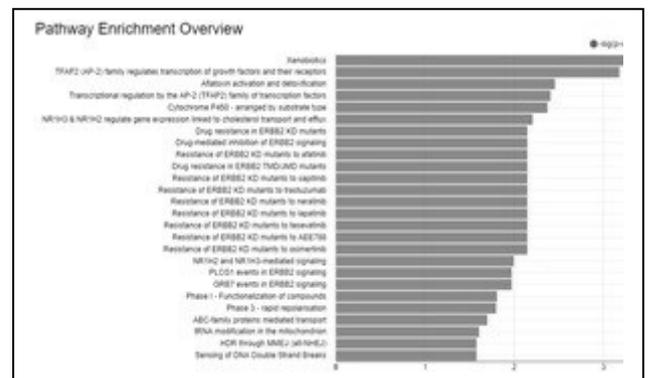

*Figure 4: Identified Pathways related to significantly expressed genes.*

### 3.2 PGM – dependence decomposition

One of the feasible PGM models to create from the observed genes' expressions is one that aims to reverse engineer the Gene Regulatory Network (GRN) regulating the genes expressed in the experiment. As a result, we needed to embed existing known biological knowledge to create the network that identifies a parent gene to a child gene (parent gene regulates the expression of a child gene). We searched GeneMania [16] for gene interactions using the most significant gene identified in the previous step. Only 19 genes were identified by GeneMania and their network was extracted as illustrated in Figure 5. An arrow connecting these genes explains a co-expression ranked by maximizing the Pearson correlation coefficient for the query gene pair. This score represents the expression level similarity across several experiments. Co-expression does not mean causality or distinguish between regulatory and regulated genes. To identify causality, more data need to be added such as sequence motif analysis, protein-protein interaction, Transcription Factors (TF) and target interactions, and methylome data such as the work presented in [7] and [8]. These data are readily available in public databases identified in these studies and can be integrated into future work.

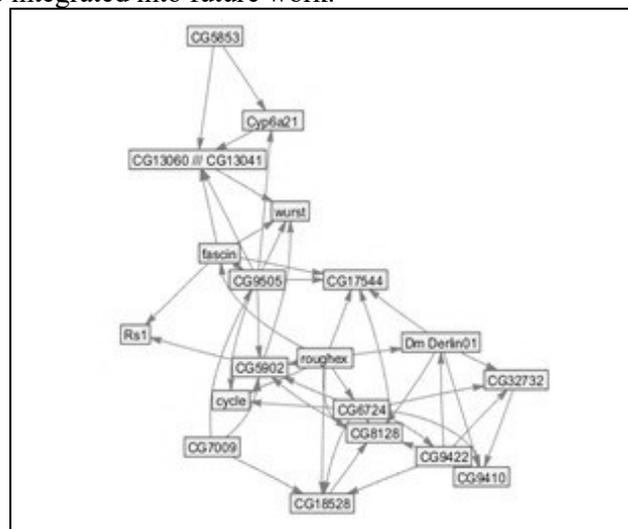

*Figure 5: Genes Interactions Network extracted from GeneMania*

One approach to infer the posterior from the data is to apply a Variational Bayesian Inference (VI) method to analytically estimate the network parameters and latent variables from the observed variables using the Expectation Maximisation (EM) algorithm. The approach estimates the parameters that fit the given data by alternating between estimating the parameters' values, and maximizing the posterior estimation (MAP) until convergence. The pre-processed 19 genes and their known co-expression were used to build a probabilistic factor graph network (FGN) as described in [17], to infer the marginal posterior distribution of the latent variables (biological function maintained by the messages communicated by the gene expressions). This method applied a discretisation step using a mixture of Gaussian distribution to model the relations between continuous observations on a gene variable and its discrete logical state and EM algorithm to infer the posterior distributions and re-estimate the mixture statistics until convergence. Then a variational algorithm known as a loopy-belief propagation (LBP) was applied to estimate the marginal posterior distributions on all gene logical variables and compared its predicted marginals with the genes observed states from the input network.

### 3.3 PGM – Bayesian Inference

Another experiment was conducted using sampling algorithms to infer the posterior. Bayesian Inference requires estimating the parameters of the distributions of the likelihood and the priors used to calculate the posterior. This is performed by iterating through the data and sampling new similarly distributed data. Sampling algorithms such as the MCMC algorithm, the Hamiltonian Monte Carlo implementation in Stan code as described in [6] was used.

Bayesian inference generally starts by identifying the parameters to infer. The parameter to infer is the posterior probability distribution of the gene expression $\theta_i$ for the i$^{th}$ gene in a given transcriptome in a given sample in an experiment such as [11]. Having the joint priors and conditional decomposition, we can apply the probability chain rule to infer $\theta_i$ as follows:

$$P(\theta_1, \theta_2, \ldots, \theta_n) = \prod_{i-1}^{n} \begin{cases} P(\theta_i) & for\ a\ parentless\ paramter \\ P(\theta_i|\ parents(\theta_i)) & for\ a\ paramter\ with\ parent(s) \end{cases}$$

The posterior distribution is conjugate, which means it is the same as the prior distribution of the gene expressions. The model in [6] uses the Binomial Distribution for one gene (and Multinomial for GWAS) for the posterior distribution. The conjugate Beta distribution for one gene (and Dirichlet Distribution for N genes) for the prior distribution since it is suitable to the random behaviour of percentages and proportions capturing the uncertainty in the expression levels of the various genes.

The next step is to identify the available dataset as follows: Gene expressions, number of genes in the model (19 the identified significant genes), number of samples (two conditions – control vs SMN mutant), number of tissues (three: Whole larvae, Brain and Muscle Tissues).

The identification of the priors requires subjective evaluation of gene expression. This case study adopts the model in [6], by using uninformative distribution such as Beta that can accommodate various distributions by estimating both its parameters from the given data:

$$Beta\ (\alpha, \beta) = \int_0^1 \theta^{\alpha-1}.(1-\theta)^{\beta-1}d\theta$$

For a single gene: $P(\theta_i) = Beta\ (\alpha, \beta)$
For all the genes at one closed-form: $P(\vec{\theta}) = Dirichlet\ (\vec{\alpha})$

Then we need to identify the likelihood function based on the identified distributions:

$$P(k_i|n, \theta_i) = Binomial\ (k_i|n, \theta_i) = \binom{k_i}{n} \theta_i^{k_i} (1-\theta_i)^{n-k_i}$$

$$P(\vec{k}|\vec{\theta}) = Multinomial\ (\vec{k}|\vec{\theta})$$

Finally, the posterior is Dirichlet distribution as well:

$$P(\vec{\theta}|\vec{k}) = \frac{P(\vec{k}|\vec{\theta}) \times P(\vec{\theta})}{P(\vec{k})} = Dirichlet\ (\vec{\alpha} + \vec{k})$$

However, because we are modelling the variation in gene expressions across samples in different conditions, the variability in gene counts in each condition is expected to be high. This justifies using a Negative Binomial distribution as the discrete version of Gamma distribution that is more suitable in modelling waiting times such as survival analysis. We start from an informed prior model "alpha" with a common distribution that is described by two other parameters, $\mu_\alpha$, $\sigma_\alpha$ (mu_alpha and sigma_alpha in the code). The following are the parameters to estimate simplifying them to normal distributions:

- $\vec{\mu}$ represents the gene expression across every gene in the transcriptome, corresponding to $\vec{\theta}$ in the general framework explained above.
- $\vec{\beta}$ represents the differences in the expression level between stages 1 and 2.
- hyperparameter $\vec{\sigma}$, (sigma in the code) which will describe the expected variability of the observed changes in expression between both stages.
- hyperparameter $\sigma_\beta$ will describe the expected variability of the observed changes in expression and is the expected standard deviation for $\vec{\beta}$.

## 4    Results And Discussions

The FGN as described in [17] performance is illustrated in Figure 6. It shows that the algorithm converged after six iterations only by plotting the correlation coefficient against the increasing iterations. This step produced joint probability for each gene on/off status. The result for the SMA network is illustrated in Figure 7 showing the predicted vs actual states with Correlation coefficient r=0.85 and P-value 9.28e-12. In this model, we used two-quantization levels with values 0,1 states. However, genes may not be on/off only, they can also be in various states such as severely under/overexpressed, moderately under/overexpressed, normal or any other number of states as required by the analyst. This is interpreted as each state (i.e., Gaussian component) may correspond to a different range of gene-expression levels for different genes, defined by the estimated parameters of the Gaussian mixture model (GMM).

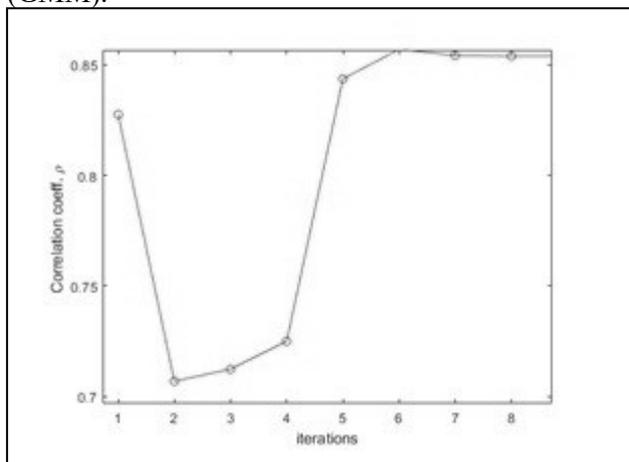

*Figure 6: Pearson correlation plots of LBP message-passing convergence with increasing iteration for 2-states discretization levels in SMA response network*

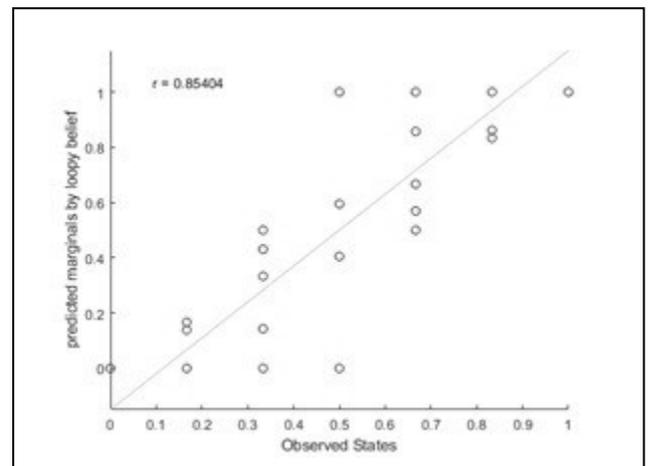

*Figure 7: Pearson correlation plots between proportions of observed states and FGN inferred marginal posteriors for SMA response network: 2-states discretization, P-value 9.2836e-12 Correlation coefficient r is given in the plot*

Table 1 lists the alpha parameters convergence values and Table 2 lists the beta parameters convergence values after the MCMC sampling using the given Bayesian model explained in the previous section and adopted in [6]. Further performance evaluations could have evaluated the error if we had a third stage gene expressions' counts or subjective evaluations by a medical expert. Figures 8 to 12 illustrate the same results showing median only and distributions for every parameter for visualisation of the performance of each of the 19 genes.

**Table 1: Posterior Statistics for alpha parameters**

| Gene | Variable/Symbol | | Median | Mean | StdDev |
|---|---|---|---|---|---|
| | mu_alpha | $\mu_\alpha$ | -0.4192 | -0.4193 | 0.0294 |
| | sigma_alpha | $\sigma_\alpha$ | 0.1071 | 0.1094 | 0.0206 |
| CG13060/ CG13041 | alpha[1] | $\mu_1$ | 0.4646 | -0.4635 | 0.0340 |
| CG17544 | alpha[2] | $\mu_2$ | 0.4610 | -0.4600 | 0.0305 |
| CG18528 | alpha[3] | $\mu_3$ | 0.4885 | -0.4881 | 0.0337 |
| CG32732 | alpha[4] | $\mu_4$ | 0.3697 | -0.3709 | 0.0340 |
| CG5853 | alpha[5] | $\mu_5$ | 0.2903 | -0.2919 | 0.0337 |
| CG5902 | alpha[6] | $\mu_6$ | 0.3647 | -0.3616 | 0.0353 |
| CG6724 | alpha[7] | $\mu_7$ | 0.4812 | -0.4797 | 0.0341 |
| CG7009 | alpha[8] | $\mu_8$ | 0.3815 | -0.3799 | 0.0308 |
| CG8128 | alpha[9] | $\mu_9$ | 0.4308 | -0.4336 | 0.0379 |
| CG9410 | alpha[10] | $\mu_{10}$ | 0.2364 | -0.2373 | 0.0336 |
| CG9422 | alpha[11] | $\mu_{11}$ | 0.4142 | -0.4143 | 0.0303 |
| CG9505 | alpha[12] | $\mu_{12}$ | 0.3700 | -0.3695 | 0.0298 |
| cycle | alpha[13] | $\mu_{13}$ | 0.5665 | -0.5639 | 0.0351 |
| Cyp6a21 | alpha[14] | $\mu_{14}$ | 0.2250 | -0.2255 | 0.0318 |
| Dm Derlin01 | alpha[15] | $\mu_{15}$ | 0.4493 | -0.4493 | 0.0324 |
| fascin | alpha[16] | $\mu_{16}$ | 0.6003 | -0.5989 | 0.0300 |
| roughex | alpha[17] | $\mu_{17}$ | 0.4786 | -0.4771 | 0.0323 |
| Rs1 | alpha[18] | $\mu_{18}$ | 0.4659 | -0.4666 | 0.0330 |
| wurst | alpha[19] | $\mu_{19}$ | 0.4659 | -0.4642 | 0.0309 |

**Table 2: Posterior Statistics for beta parameters**

| Gene | Variable/Symbol | | Median | Mean | StdDev |
|---|---|---|---|---|---|
| Hyperparam | sigma_beta | $\sigma_\beta$ | 0.0537 | 0.0558 | 0.0152 |
| CG13060 / CG13041 | beta[1] | $\beta_1$ | 0.0147 | -0.0144 | 0.0298 |
| CG17544 | beta[2] | $\beta_2$ | 0.0159 | -0.0153 | 0.0259 |
| CG18528 | beta[3] | $\beta_3$ | 0.0161 | -0.0160 | 0.0300 |
| CG32732 | beta[4] | $\beta_4$ | 0.0226 | -0.0215 | 0.0241 |
| CG5853 | beta[5] | $\beta_5$ | 0.0159 | -0.0167 | 0.0250 |
| CG5902 | beta[6] | $\beta_6$ | 0.0165 | -0.0162 | 0.0241 |
| CG6724 | beta[7] | $\beta_7$ | 0.0144 | -0.0143 | 0.0262 |
| CG7009 | beta[8] | $\beta_8$ | 0.0090 | -0.0088 | 0.0290 |
| CG8128 | beta[9] | $\beta_9$ | 0.0131 | -0.0148 | 0.0309 |
| CG9410 | beta[10] | $\beta_{10}$ | 0.0030 | -0.0025 | 0.0260 |
| CG9422 | beta[11] | $\beta_{11}$ | 0.0183 | -0.0188 | 0.0289 |
| CG9505 | beta[12] | $\beta_{12}$ | 0.0366 | -0.0344 | 0.0251 |
| cycle | beta[13] | $\beta_{13}$ | 0.1282 | 0.1270 | 0.0358 |
| Cyp6a21 | beta[14] | $\beta_{14}$ | 0.0026 | 0.0031 | 0.0276 |
| Dm Derlin01 | beta[15] | $\beta_{15}$ | 0.0101 | -0.0091 | 0.0264 |
| fascin | beta[16] | $\beta_{16}$ | 0.1004 | 0.0995 | 0.0326 |
| roughex | beta[17] | $\beta_{17}$ | 0.0601 | 0.0596 | 0.0296 |
| Rs1 | beta[18] | $\beta_{18}$ | 0.0389 | 0.0394 | 0.0332 |
| wurst | beta[19] | $\beta_{19}$ | 0.0039 | -0.0054 | 0.0254 |

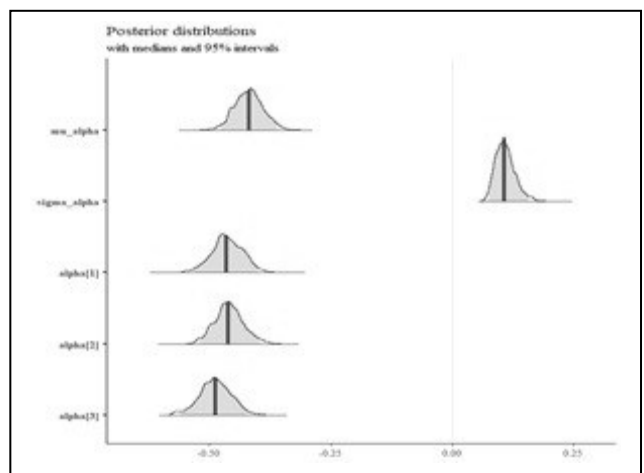

*Figure 8: Posterior Converged Values for $\mu_\alpha$, $\sigma_\alpha$, $\mu_1$, $\mu_2$, $\mu_3$ parameters.*

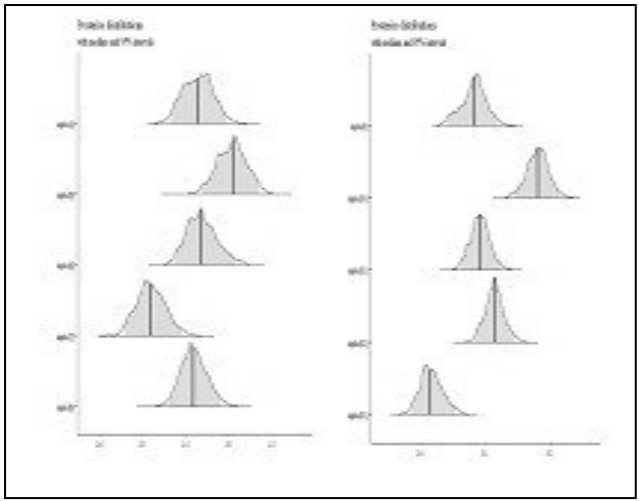

*Figure 9: Posterior Converged Values for $\mu_4$, $\mu_5$, $\mu_6$, $\mu_7$, $\mu_8$, $\mu_9$, $\mu_{10}$, $\mu_{11}$, $\mu_{12}$, $\mu_{13}$ parameters.*

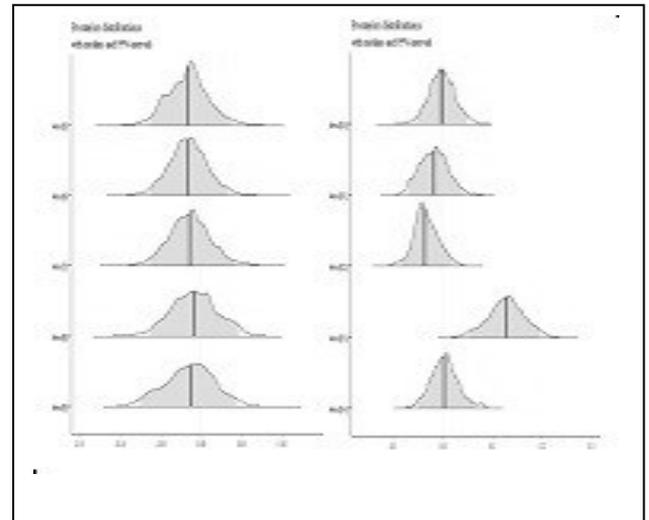

*Figure 11: Posterior Converged Values for $\beta_5$, $\beta_6$, $\beta_7$, $\beta_8$, $\beta_9$, $\beta_{10}$, $\beta_{11}$, $\beta_{12}$, $\beta_{13}$, $\beta_{14}$ parameters.*

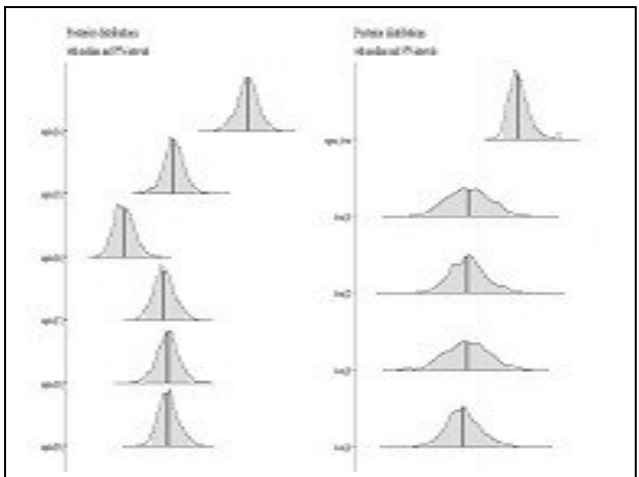

*Figure 10: Posterior Converged Values for $\mu_{14}$, $\mu_{15}$, $\mu_{16}$, $\mu_{17}$, $\mu_{18}$, $\mu_{19}$, $\sigma_\beta$, $\beta_1$, $\beta_2$, $\beta_3$, $\beta_4$ parameters.*

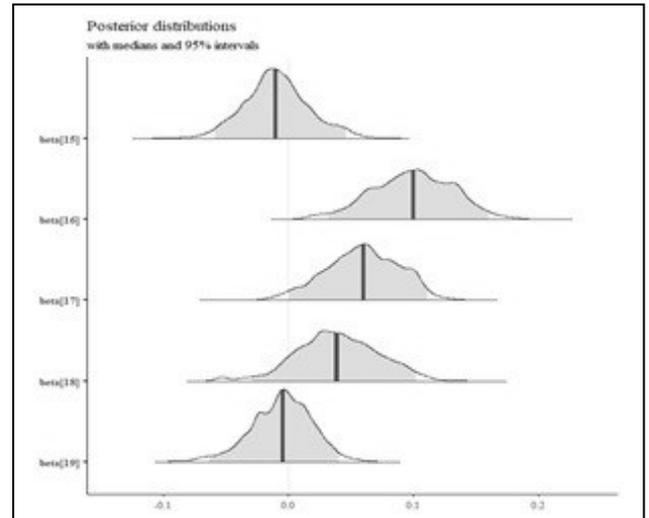

*Figure 12: Posterior Converged Values for $\beta_{15}$, $\beta_{16}$, $\beta_{17}$, $\beta_{18}$, $\beta_{19}$ parameters*

## 5   Conclusion

This experiment shows how to semi-automate the pre-processing of the GWAS generated gene expressions data to model a PGM to reverse engineer the GRN co-expression patterns in two different development stages of a disease. The previous PGM SMA model was hand modelled by experts in [10] without publishing the details of the model and evaluate its performance. The PGM we automatically created from GWAS, is enriched with prior biological knowledge extracted from public databases that summarize biological experiments. Pathways linked to the most significantly expressed genes are identified. Adopting the methods described in [17] for the SMA dataset shows how a variational Bayesian method such as the loopy-belief propagation (LBP) is applied to analytically infer the posterior. The work in [5] was adopted to use an iterative sampling method such as MCMC using the Hamiltonian Monte Carlo algorithm to infer the parameters' estimates of the posterior distribution of each gene and can be used for prediction. MCMC approaches are computationally expensive but have no bias and produce more accurate results than VI algorithms. VI approaches introduce a bias but performs a reasonable optimisation process suitable to very large-scale problems.

Both methods require some performance evaluations, whether computationally by having values for the third stage counts, or medically by checking if these results are sound or not. The experiment will benefit from bigger datasets, preferably on several stages/conditions of a disease progression or controlled drug administration experiments. Having a posterior distribution for gene expressions over two stages of disease development can be used to infer causality or predictions. The accuracy of these causalities or predictions will increase with having bigger datasets, over many stages of disease development or drug administration or condition to be investigated. The work shows how data science using open-source software and public databases can model biological and medical experiments data for further analysis and prediction.

Future work can aim to further connect with drugs databases such as ChEMBL [18] for drugs that are known to target these genes expressions and pathways. Another suitable type of PGM is the Higher-Order Dynamic Bayesian Network (HO-DBN) and the implementation discussed in [19] to benefit from the current SMA gene expressions that were collected in two different time steps. The current limitation is that the minimum required for this method is three-time steps. This is not available in the SMA GWAS experiment in [11].

Further future work can proceed to add various hierarchies for the linked pathways, CRMs, TFs such as explained in [7], or the cause/effect relationships between genes such as is explained in [8]. The uncertainty of the chosen priors is always overridden by feeding the model with more data. If public hospitals worldwide publish their gene expressions data for various case/control studies at various disease development stages, or various controlled drug administration stages, the models like these generated in the experiments we conducted, or the ones identified in the future work, will achieve higher predictive accuracy for further investigation and data integration.